\documentclass[conference]{IEEEtran}
\IEEEoverridecommandlockouts
\usepackage{cite}
\usepackage{amsmath,amssymb,amsfonts}
\usepackage{algorithmic}
\usepackage{graphicx}
\usepackage{textcomp}
\usepackage{xcolor}
\def\BibTeX{{\rm B\kern-.05em{\sc i\kern-.025em b}\kern-.08em
    T\kern-.1667em\lower.7ex\hbox{E}\kern-.125emX}}
\begin{document}

\title{Leveraging Context-awareness to Better Support the IoT Cloud-Edge Continuum
}

\author{\IEEEauthorblockN{Liliana I. Carvalho}
\IEEEauthorblockA{\textit{COPELABS, Lusófona University}\\
Lisbon, Portugal \\
liliana.inocencio@ulusofona.pt}
\and
\IEEEauthorblockN{ Daniel Maniglia A. da Silva}
\IEEEauthorblockA{\textit{COPELABS, Lusófona University}\\
Lisbon, Portugal \\
daniel.maniglia@hotmail.com}
\and
\IEEEauthorblockN{ Rute Carvalho Sofia}
\IEEEauthorblockA{
\textit{fortiss}, 
Munich, Germany \\
\textit{Lusófona University, Lisbon, Portugal}\\
sofia@fortiss.org}
}

\maketitle

\begin{abstract}
Novel Internet of Things (IoT) requirements derived from a broader interconnection of heterogeneous devices have pushed the horizons of Cloud computing and are giving rise to a wider decentralisation of applications and data centers. An answer to the underlying network concerns, such as the need to lower the resulting latency due to heavy computation needs, or safety aspects, gave rise to Edge/Fog computing, where IoT functionality can be also supported closer to data sources. While it is today feasible to perform some IoT functionality on the Edge, the orchestration of operations between Edge and Cloud requires an automated support, where context-awareness plays a key role in assisting the network in deciding when and where to store data and to perform computation. This work is focused on the application of context-awareness to support a smoother operation of the Edge to Cloud operation, aiming at lowering latency, in particular when real-time or close-to-real-time data exchange is present.  \end{abstract}

\section{Introduction}
IoT scenarios are today being supported by computational paradigms, such as Cloud computing or Edge/Fog computing \cite{b1}. \emph{Fog Computing} is a set of paradigms that assist computation, networking, and storage between the edges of the network and the Cloud. Fog Computing therefore extends the Cloud capabilities to the edges of the network and closer to data sources, thereby providing the means to  support better real-time data processing. In Fog Computing, resources are dynamically distributed across the Cloud and Edge elements based on, for instance, \emph{Quality of Service (QoS)} requirements \cite{b2}, specific network policies, and sometimes: application requirements. 
However, there are today new IoT applications emerging and in particular computational intensive applications such as Augmented Reality (AR), which are also often distributed (e.g., Mobile Crowd Sensing platforms). For such applications, there are challenges to be met by the underlying networking architectures, for instance: handling mobility of specific micro-services and devices; constrained resources, etc. Such challenges are not yet fully envisioned in Edge/Cloud architectures. 

A way to address the issues concerning a better support of IoT operations across the Edge and Cloud is to consider context-awareness to assist a decision on where and how to store and to compute data. Context-awareness usually implies considering aspects such as node features (e.g., resource usage, such as CPU; but also soft features such as betweenness and availability); network features (e.g., congestion level, available bandwidth). In the case of our work, it also means considering specific application requirements (e.g., low latency) and above all: integrating into the support of Edge to Cloud operations \emph{social behaviour} aspects, e.g., roaming habits of users \cite{b3}.

The focus of our work is therefore on contributing to a better support of IoT, via the development of context-aware network algorithms and mechanisms that are expected to result in reducing latency, while better supporting heterogeneous IoT devices, including the potential mobility of Edge devices.

\section{Research Problem and Related Work}
IoT environments are becoming more and more people-centric. For instance, Smart Cities today integrate large-scale pervasive sensing environments which are based on sets of static and mobile sensors and actuators, including personal mobile devices. Also, smart health integrates IoT systems that are also often based on devices that are carried by the user such as smartphones, smart watches, wristbands, or devices that are controlled by the user, e.g., specific IoT sets. Today, all of the data is being sent to the Cloud, bringing in several challenges \cite{b4}. Firstly, there is an associated delay \cite{b5}, often not compatible with advanced IoT applications. Secondly, sensitive data is being sent to the Cloud, thus requiring additional strong security measures \cite{b6}. Thirdly, decentralised services have to depend on central control mechanisms, thus increasing computational complexity.

However, to solve these issues, research is addressing possibilities to bring computation, storage, and data processing as close as possible to data sources. Some, if not most of the data collected can be locally handled instead of being sent to the Cloud in an indiscriminate way. This may assist aspects such as data and user privacy. Local computation can also assist in reducing overhead of these systems, as aspects such as calibration may be partially handled closer to the end-user.

The possibility to handle data closer to the source is being addressed in several research fronts, as detailed in the next sub-sections, by bring computation, storage, and data processing to the Edge. 

Moreover, and in addition to provide a more flexible network control, as occurs in Fog architectures, our proposal is to integrate context-awareness into the network layers, both in order to assist in a better decision on whether data should go the Cloud or stay in the Edge, as well as to better assist decentralised IoT services, such as, for instance, data analytics, Mobile Crowd Sensing.

\begin{figure*}[h]
    \centering
    \includegraphics[width=0.7\textwidth]{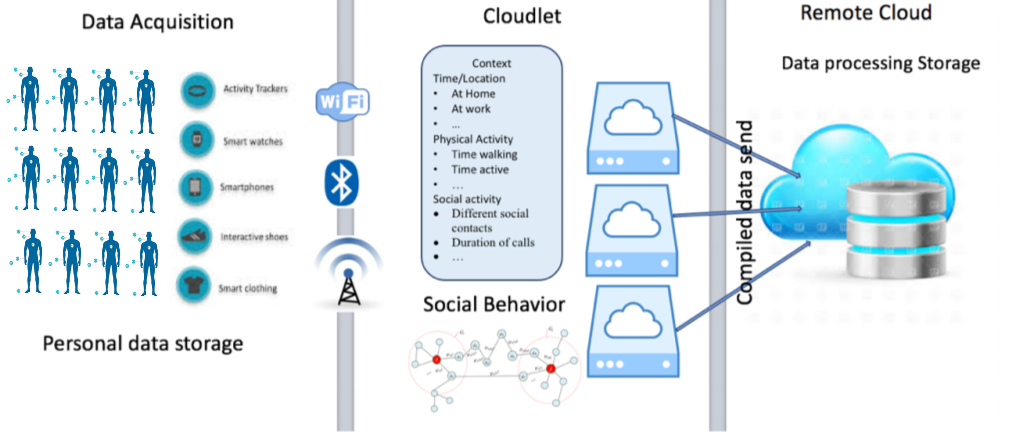}
    \caption{Mobile Crowd-based Sensing Scenario}
    \label{figure1}
\end{figure*}

\begin{figure*}[h]
    \centering
    \includegraphics[width=0.7\textwidth]{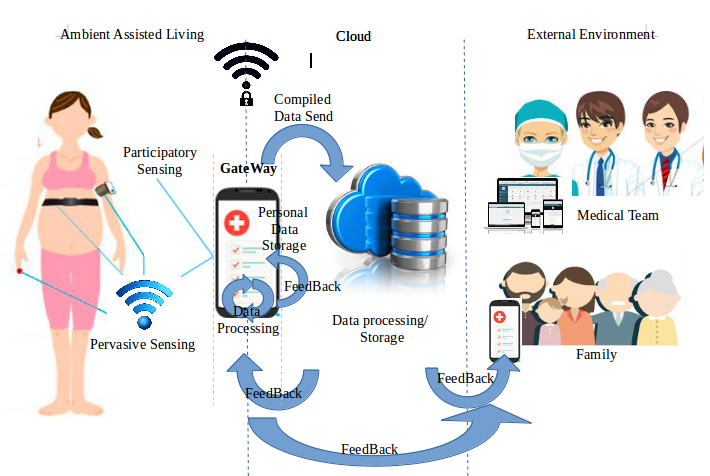}
    \caption{Smart Health use-case, pregnancy support on-the-go}
    \label{figure2}
\end{figure*}

\subsection{Edge Computing}

According to Bonomi et al., Edge computing is a networking paradigm that extends Cloud Computing to the edges of the network, where the Edge can be placed in different locations, e.g., an Evolved NodeB, an Access Router, an IoT gateway, or even: a smartphone. Edge computing brings in new opportunities to explore a new generation of IoT applications and services \cite{b1}. By assisting the placement of storage, data processing and analytics, and computation closer to the data sources, Edge Computing brings in benefits in terms of latency and energy consumption \cite{b7}. The main goal of Edge Computing is to extend the Cloud capabilities to the edges of the network, thereby supporting real-time data processing and latency sensitive applications \cite{b8, b9, b10, b11}. For IoT, and due to aspects such as security (e.g., the need to have in-plant security and resilient communication), large distances, as well as large sets of frequent data can lead to an insufficiency of the Cloud in satisfying QoS requirements (e.g., low latency) of different IoT applications. Hence, Fog computing aims to overcome some limitations of Cloud-centric IoT-models by taking advantage of edge resources \cite{b12}.

Some prior work focuses on definitions of Edge Computing for IoT. Cheng et al. developed FogFlow \cite{b13}, a Edge/Fog Computing software platform that provides a programming model for IoT-based applications for smart cities over cloud and edges.  Saurez et al. developed Foglets \cite{b14} a programming infrastructure for large scale situation awareness application, focused on the optimisation of task deployment over Fog with the purpose of save bandwidth and reduce latency. Hong et al. \cite{b15} proposed Mobile Fog a high-level situation awareness programming model in fog computing platform for large-scale and latency sensitive applications. In a study about the placement of IoT services from Skarlat et al. \cite{b16}, the authors proposed a framework for fog applications consisting of many services, taking in consideration the QoS requirements of the resources.

\subsection{Context-Awareness}
Adequate network support to reduce resource consumption in large-scale pervasive IoT will benefit from integrating context-awareness. In the context of IoT, where smart applications are used to collect and to exchange different types of data, Sholl et al. propose a smart city architecture that harnesses the power of semantic computing that allows machines and people to understand the relationships among data in a context-aware manner, to extract knowledge \cite{b17}.
Context-awareness becomes relevant in such environments, to assist a better management of infrastructures. For instance, Choi et al. propose a software architecture to assist efficient middleware deployment in Smart Cities, by relying on semantic computing \cite{b18}.
There are several benefits derived from considering context-awareness into Fog Computing mechanisms \cite{b19}: reduction of network congestion; efficient resource and provision of spontaneous services; a better system adaptability to surrounding context.

Context-awareness is also highly relevant to assist better data mining and classification, as debated in the context of vehicular networks by Ruta et al. \cite{b20}. Device usage data is also being used to improve the efficiency in Edge Computing.

\section{Research Relevancy}
IoT environments are becoming increasingly people-centric, as already explained. Moreover, some, if not most of the transmitted data can be locally handled, e.g., in the Edge, in Cloudlets. Bringing data closer to the sources can assist in reducing the overall latency and control overhead. However, this also requires a different, more interactive type of data exchange to scale well in heterogeneous, mobile environments. 

Based on the state of the art, it is possible to state that context-awareness can assist the network in making decisions that improve the network operation in real time system. Identification of the advantages, disadvantages and challenges of such process is an aspect that will be worked upon. The Edge Computing architecture seams to decrease network latency. Today, context data is until limited and people-centric data context are not being considered. 

\section{Research Approach and Key Contributions}\label{AA}

The research approach considers two key aspects: i) applying context-awareness to perform edge selection; ii) applying context-awareness to better adapt the network to the requirements of large-scale sensing environments, for instance, Mobile Crowd Sensing platforms.

Today, this notion of context is associated with the network and node workload, as well as with node location. However, other context data can be used to improve the network management or Edge selection. For instance, one can consider application requirements (e.g., memory, CPU consumption, duration of use); node availability in regards to data processing; system history of usage and end-user behaviour (e.g., mobility patterns).

The work being developed considers two applicability scenarios being evaluated in FogFlow \cite{b13}: a Mobile Crowd Sensing scenario; ii) a people-centric IoT (Smart Health) scenario.  Both scenarios are being deployed and evaluated in terms of latency and energy consumption. 

 FogFlow \cite{b13} is a framework which provides a standards-based programming model for IoT services that run over cloud and edges. The FogFlow framework enables programming of IoT services and it supports standard interfaces for contextual data transfers across services. 

Figure \ref{figure1} presents the first scenario, the Mobile Crowd-based Sensing use case. This scenario is focused on the reduction of end-to-end latency and energy consumption reduction in large-scale sensing environments via the development of mechanisms that can: i) leverage networking architectures and in particular, Fog architectures, so that they better adjust to large-scale distributed sensing platforms; assuming highly mobile environments, and volatile data; ii) assist edge based classification, providing the means for hierarchical classification, and integrating  social behavior inference aspects, e.g., roaming habits.

The second scenario (Smart Health IoT), ilustrated in \ref{figure2}, focuses on the deployment of novel context-aware edge selection algorithms. The health care scenario brings characteristics of real-time systems, where latency is always a concern. In addition to the need for collaboration (feedback) between the systems and the users involved in patient care, such as family members and/or the medical team. The proposed algorithms are being evaluated against existing edge selection and edge data placement algorithms. 

Today, there is related work focused on some context information associated with location or device capacity and usage. Our ongoing work contributes by considering other context-awareness indicators such as user activity, user connection history, or displacement patterns, for example.


\begin{thebibliography}{00}

\bibitem{b1} F. Bonomi, R. Milito, J. Zhu, and S. Addepalli, “Fog computing and its role in the internet of things,” in Proceedings of the first edition of the MCC workshop on Mobile cloud computing. ACM, pp. 13–16, 2012.

\bibitem{b2} D. Sarddar, S. Barman,P. Sen, and R. Pandit, “Refinement of Resource Management in Fog Computing Aspect of QoS,“ International Journal of grid and Distributed, vol. 11. number 5, pp.29--44, 2018.

\bibitem{b3} C. Li, K., B. Di Martino, T. Yang, L., and  Q. Zhang, "Smart data: State-of-the-art perspectives in computing and applications," CRC Press, 2019.

\bibitem{b4} W. Masri, I. Al Ridhawi, N. Mostafa and P. Pourghomi, "Minimizing delay in IoT systems through collaborative fog-to-fog (F2F) communication," 2017 Ninth International Conference on Ubiquitous and Future Networks (ICUFN), Milan, pp. 1005-1010, 2017.

\bibitem{b5} S. Mubeen, P. Nikolaidis, A. Didic, H. Pei-Breivold , K. Sandstrom, and M. Behnam, “Delay mitigation in offloaded cloud controllers in industrial IoT,” IEEE Access, vol. 5, pp. 4418–4430, 2017.

\bibitem{b6} S. T. Mohamad and K. N. Yasen. “Homomorphic encryption implementation to ensure data security in cloud computing.” Journal of Theoretical and Applied Information Technology, vol. 96, no. 7, pp. 1826-1836, 2018


\bibitem{b7} J. Gedeon, F. Brandherm, R. Egert, T. Grube, and M. Muhlhauser, “What the fog? edge computing revisited: Promises, applications and future challenges,” IEEE Access, vol. 7, pp. 152 847–152 878, 2019.

\bibitem{b8} D. Sarddar, S. Barman, P. Sen, and R. Pandit, “Refinement of resource management in fog computing aspect of qos,” International Journal of Grid and Distributed Computing, vol. 11, no. 5, pp. 29–44, 2018.

\bibitem{b9} M. Iorga, L. Feldman, R. Barton, M. J. Martin, N. S. Goren, and C. Mahmoudi, “Fog computing conceptual model,” NIST: National Institute of Standards and Technology, 2018.

\bibitem{b10} T. Choudhari, M. Moh, and T.-S. Moh, “Prioritized task scheduling in fog computing,” in Proceedings of the ACMSE 2018 Conference. ACM, p. 22, 2018.

\bibitem{b11} P. Maiti, J. Shukla, B. Sahoo, and A. K. Turuk, “Qos-aware fog nodes placement,” in 2018 4th International Conference on Recent Advances in Information Technology (RAIT). IEEE, pp. 1–6, 2018.

\bibitem{b12} M. Aazam, S. Zeadally, and K. A. Harras, “Deploying fog computing in industrial internet of things and industry 4.0,” IEEE Transactions on Industrial Informatics, vol. 14, no. 10, pp. 4674–4682, 2018.

\bibitem{b13} B. Cheng, G. Solmaz, F. Cirillo, E. Kovacs, K. Terasawa, and A. Kitazawa, “FogFlow: Easy Programming of IoT Services Over Cloud and Edges for Smart Cities,” IEEE Internet of Things Journal, vol. 5, no. 2, pp. 696–707, 2018.

\bibitem{b14} E. Saurez, K. Hong, D. Lillethun, U. Ramachandran, and B. Ottenwälder, “Incremental deployment and migration of geo-distributed situation awareness applications in the fog,” in Proc. 10th ACM Int. Conf. Distrib. Event Based Syst., Irvine, CA, USA, pp. 258–269, 2016.


\bibitem{b15} K. Hong, D. Lillethun, B. Ottenwalder, and B. Koldehofe, “Mobile Fog : A Programming Model for Large Scale Applications on the Internet of Things,” in The second ACM SIGCOMM f (MCC ’13), pp. 15–20, 2013.

\bibitem{b16} O. Skarlat, M. Nardelli, S. Schulte, and S. Dustdar, “Towards QoS-Aware Fog Service Placement,” Proceedings - 2017 IEEE 1st International Conference on Fog and Edge Computing, ICFEC 2017, pp. 89–96, 2017

\bibitem{b17} S. Sholla, R. Naaz, and M. A. Chishti, “Semantic smart city: Context aware application architecture,” in 2018 Second International Conference on Electronics, Communication and Aerospace Technology (ICECA). IEEE, pp. 721–724, 2018.

\bibitem{b18} C. Choi, C. Esposito, H. Wang, Z. Liu, and J. Choi, “Intelligent power equipment management based on distributed context-aware inference in smart cities,” IEEE Communications Magazine, vol. 56, no. 7, pp. 212– 217, 2018.

\bibitem{b19} F. Hussain, Internet of Things; Building Blocks and Business Modles. Springer, 2017.

\bibitem{b20} M. Ruta, F. Scioscia, F. Gramegna, S. Ieva, E. Di Sciascio, and R. P. De Vera, “A knowledge fusion approach for context awareness in vehicular networks,” IEEE Internet of Things Journal, vol. 5, no. 4, pp. 2407–2419, 2018.













\end{thebibliography}
\end{document}